\documentclass[epj]{svjour}
\usepackage{graphicx}
\usepackage{hyperref}
\usepackage{multirow}
\usepackage{float}
\usepackage{amsmath,amsfonts,amssymb}
\usepackage{subcaption}
\usepackage{tikz}
\begin{document}
\title{Characterisation of the waveplate associated to layers in interferential mirrors}

\author{J. Agil\inst{1}\inst{*} B. Letourneur\inst{2} \and S. George\inst{1} \and R. Battesti\inst{3} \and C. Rizzo\inst{3}}
\institute{ 
\inst{1} LNCMI, CNRS, UT3, UGA, INSA-T, EMFL, F-31400 Toulouse Cedex, France\\
\inst{2} SAFRAN ELECTRONICS \& DEFENSE, \'{E}tablissement de Montlu{\c c}on, BP 3247 - 03106 MONTLU{\c C}ON CEDEX\\
\inst{3} Universit\'{e} Toulouse III - Paul Sabatier, LNCMI, CNRS, UGA, INSA-T, EMFL, F-31400 Toulouse Cedex, France\\
\inst{*} Corresponding author: jonathan.agil@lncmi.cnrs.fr
}

\date{Received: date / Revised version: date}
%
\abstract{In this paper, first we present a review of experimental data corresponding to phase retardation per reflection of interferential mirrors. Then, we report our new measurements on both commercial and tailor-made mirrors. To be able to measure the phase retardation as a function of the number of layers, additional pairs of layers are deposited on some of the mirrors. The results obtained with this special set of mirrors allow us to fully characterise the waveplate associated with the additional pairs of layers. We finally implemented a computational study whose results are compared with the experimental ones. Thanks to the additional layers, we have achieved reflectivity never measured before at $\lambda=1064$~nm, with an associated finesse of $\mathcal{F}=895~000$.
} 
\maketitle
\section{Introduction}

Since the 80ies of the last century, it is known that interferential mirrors act as waveplates on the light reflected even at 0$^\circ$ degree incidence (see~\cite{Yoshino1979} and references within). Interferential mirrors are essentially a stack of dielectric layers of high and low refractive indices~\cite{Born1983}. Mirror reflectivity $R_{\mathrm{th}}$ depends on the number $N$ of high and low refractive indices layers, respectively $n_H$ and $n_L$. Following~\cite{Born1983}, one can write
\begin{equation}\label{eq:BruitOpt:R}
R_{\mathrm{th}}=\left(\frac{1-\left(\dfrac{n_H}{n_0}\right)\left(\dfrac{n_H}{n_s}\right)\left(\dfrac{n_H}{n_L}\right)^{2N}}{1+\left(\dfrac{n_H}{n_0}\right)\left(\dfrac{n_H}{n_s}\right)\left(\dfrac{n_H}{n_L}\right)^{2N}}\right)^2,
\end{equation} 
with $n_s$ the index of refraction of the substrate and $n_0=1$ the one of the medium above the stack. The Eq.~\ref{eq:BruitOpt:R} implies that an additional layer of refractive index $n_H$ is added at the top of the stack totalling $2N+1$ layers. 

Understanding the origin of such a birefringence, and therefore acquiring the ability of controlling it during the fabrication process is of general interest as in the large use of interferential mirrors in metrology (see \textit{e.g.}~\cite{Millo2009}) or in fundamental physics (see \textit{e.g.}~\cite{Pold2020}). Interferential mirrors birefringence is a manifestation of the stress induced during layer deposition, and advances in optical film and coatings would also profit to an even larger range of activities (see \emph{e.g.}~\cite{Martinu2000}).

A first review of the reported measurements of interferential mirrors phase retardation per reflection can be found in~\cite{Bielsa2009}. Authors of~\cite{Bielsa2009} infer from their study that mirror phase retardation is smaller in mirrors with higher reflectivities and that data seem to indicate the existence of a trend corresponding to a phase retardation proportional to experimental $1-R$, denoted the ``B-et-al'' trend in the following. Their computational study also shows that such a trend can be simply explained by the presence of only one birefringent layer in contact with the substrate. 

Since then new data have been published and almost all of these data seem to agree with the B-et-al trend.  As already discussed in~\cite{Bielsa2009} itself, the problem inherent in the B-et-al trend is that it has been work out of a set of data consisting of mirrors of different reflectivities but also coming form different companies that may not use exactly the same fabrication processes. The suggestion in~\cite{Bielsa2009} to overcome this problem was to measure mirror phase retardation of mirrors made by exactly the same procedure, just changing the number of layers. These has been realised and reported in~\cite{Xiao2019} by Xiao~\emph{et~al.} where the phase retardation of three mirrors built by superimposing 15, 17 and 22 pairs of layers has been measured. Their results, that we detailed in the following, fully disagree with the B-et-al trend. 

In the same paper~\cite{Xiao2019}, authors report the measurement of phase retardation as a function of the impact point on the mirror surface, confirming what has been shown in 1993~\cite{Micossi1993} on a lower reflectivity and higher diameter mirror, \emph{i.e.} that the phase retardation changes from point to point not randomly.

Let us note also that the problem of stress in thin film coatings has become more and more topical in recent years both from a general point of view, as detailed in the review paper by Abadias~\emph{et~al.}~\cite{Abadias2018}, and as far as optical elements are concerned as detailed in the review article by Wei~\emph{et~al.}~\cite{Wei2021}.

Whatever is the origin of such an intrinsic mirror phase retardation, it is also a source of noise in precision optics experiments like the attempts to measure the vacuum magnetic birefringence (VMB) (see \emph{e.g.}~\cite{Ejlli2020,Beard2021,Fan2017}) or gravitational wave (GW) detectors~\cite{Kryhin2023}.

In the VMB framework, phase retardation looks as the limiting one~\cite{Ejlli2020,Agil2022}. PVLAS group claims that the origin of this noise is thermal~\cite{Ejlli2020} as in~\cite{Kryhin2023}, while the BMV group claims  that the origin is the pointing instability on the mirror surface~\cite{Hartman2017} that, as show in~\cite{Xiao2019,Micossi1993}, does not present an uniform phase retardation.

In this paper, we first upgrade the review of~\cite{Bielsa2009} by introducing data that have been missed in 2009 and data reported since. Then, we report our new measurements on commercial mirrors manufactured by FiveNine Optics, and on mirrors fabricated by Safran Electronics \& Defense in the framework of a research collaboration. The goal of this collaboration, in the framework of the BMV project, is to find the origin of the birefringence of high reflectivity mirrors in order to control it and extinguish it. On some of the mirrors fabricated by Safran two additional pairs of layers has been deposited after a first measurement, to be able to measure the phase retardation as a function  of the number of layers. The results obtained with this special set of mirrors allow us to fully characterise the waveplate associated with the additional pairs of layers. Our results are not in agreement with the assumptions that are at the base of the B-et-al trend for the simple reason that we prove that a single pair of layer is birefringent. To try to conciliate Xiao~\textit{et~al.} results, ours and the fact that high reflectivity seems to be always associated with low phase retardation we implement a computational study based on the one of~\cite{Bielsa2009}. We consequently compare our computational results, Xiao~\textit{et~al.} ones, and ours. Let us finally note that thanks to the additionnal layers, Safran mirrors have achieved reflectivities never measured before at $\lambda=1064$~nm.

\section{Review of existing data}

In table~\ref{tab:BirMir} we list the measured reflectivity $R$, the corresponding finesse $\mathcal{F}$~\cite{Born1983}, phase retardation $\delta$, number of mirrors $N_m$ of the measured set, and wavelength $\lambda$ of published dielectric interferential mirrors.

\begin{table}
\centering
\resizebox{\linewidth}{!}{
\begin{tabular}{|c|c|c|c|c|c|}
 \hline
 \rule{0pt}{2.5ex} Ref.  & $R$ & $\mathcal{F}\times 10^{-3}$ & $\delta$ (rad) & $N_{m}$ & $\lambda$ (nm)\\
 \hline
  \rule{0pt}{2.5ex}\cite{Yoshino1979} & 0.99 & 0.3 & $2.3\times 10^{-3}$ & 1 & 633\\
  \hline 
 \multirow{2}{*}{\cite{Bouchiat1982}$^\dagger$}  & \multirow{2}{*}{0.998} & \multirow{2}{*}{1.6} & \rule{0pt}{2.5ex} $(3\pm 1)\times 10^{-4}$ & 17 & \multirow{2}{*}{0.5} \\
 \cline{4-5}
 & & & \rule{0pt}{2.5ex} $<10^{-6}$ & 2 &  \\
 \hline
 \rule{0pt}{2.5ex}\cite{Carusotto1989}$^\dagger$ & 0.996 & 0.8 & $(12.5\pm 9.5)\times 10^{-5}$ & 5 & 514\\ 
 \hline
  \rule{0pt}{2.5ex}\cite{Micossi1993}$^\dagger$ & 0.9983 & 1.8 & $(4.6\pm 1.6)\times 10^{-4}$ & 1 & 633\\ 
 \hline
   \rule{0pt}{2.5ex}\cite{Jacob1995}$^\dagger$ & 0.999524 & 6.6 & $(2.7\pm 1.7)\times 10^{-6}$ & 2 & 633\\ 
 \hline
    \rule{0pt}{2.5ex}\cite{Ni1996}$^\dagger$ & 0.9895 & 0.3 & $1.2\times 10^{-3}$ & 1 & 633\\ 
 \hline
  \rule{0pt}{2.5ex}\cite{Wood1996}$^\dagger$ & 0.999975 & 126 & $3\times 10^{-6}$ & 1 & 540\\ 
 \hline
  \rule{0pt}{2.5ex}\cite{Moriwaki1997}$^\dagger$ & 0.9988 & 2.6 & $(7.3\pm 3.1)\times 10^{-4}$ & 2 & 633\\ 
 \hline
\multirow{2}{*}{\cite{Brandi1997}$^\dagger$} & \multirow{2}{*}{0.999969} & \multirow{2}{*}{101} & \rule{0pt}{2.5ex} $(15.7\pm 8.3)\times 10^{-7}$ & 3 & \multirow{2}{*}{1064} \\
 \cline{4-5}
 &  & & \rule{0pt}{2.5ex} $<10^{-7}$ & 1 &  \\
 \hline
  \rule{0pt}{2.5ex}\cite{Hall2000}$^\dagger$ & 0.999923 & 41 & $1.8\times 10^{-6}$ & 1 & 633\\ 
 \hline
 \rule{0pt}{2.5ex}\cite{Lee2000} & 0.999963 & 85 & $(3.0\pm 0.1)\times 10^{-6}$ & 2 & 575\\ 
 \hline
 \rule{0pt}{2.5ex}\cite{Morville2002} & 0.99994 & 52 & $(5.2\pm 0.4)\times 10^{-6}$ & 2 & 1312\\ 
 \hline
 \rule{0pt}{2.5ex}\cite{Huang2008} & 0.999991 & 349 & $(5.8\pm 4.2)\times 10^{-7}$ & 2 & 1652\\ 
 \hline
\multirow{2}{*}{\cite{Bielsa2009}$^\dagger$} & 0.999396 & 5.2 & \rule{0pt}{2.5ex} $(4.6\pm 0.3)\times 10^{-4}$ & 2 & \multirow{2}{*}{1064} \\
 \cline{2-5}
 & 0.999972 & 112 & \rule{0pt}{2.5ex} $(4.5\pm 0.2)\times 10^{-6}$ & 3 &  \\
 \hline
  \rule{0pt}{2.5ex}\cite{Durand2009} & 0.999987 & 242 & $(9.6\pm 0.4)\times 10^{-7}$ & 2 & 413\\ 
  \hline
  \rule{0pt}{2.5ex}\cite{Cadene2015} & 0.9999935 & 483 & $(7.1\pm 0.1)\times 10^{-7}$ & 2 & 1064\\ 
 \hline
\rule{0pt}{2.5ex}\cite{DellaValle2016} & 0.9999959 & 766 & $(2.15\pm 0.07)\times 10^{-6}$ & 2 & 1064\\ 
 \hline   
  \rule{0pt}{2.5ex}\cite{Rivere2017} & 0.999991 & 349 & $(2.96\pm 0.02)\times 10^{-6}$ & 2 & 1064\\ 
 \hline   
  \rule{0pt}{2.5ex}\cite{Kamioka2020} & 0.9999895 & 299 & $(2.85\pm 0.1)\times 10^{-6}$ & 2 & 1064\\ 
 \hline  
 
\multirow{3}{*}{\cite{Xiao2019}} & 0.999836 & 19 & \rule{0pt}{2.5ex} $(1.43\pm 0.56)\times 10^{-6}$ & 1 & \multirow{3}{*}{633} \\
 \cline{2-5}
 & 0.999850 & 21 & \rule{0pt}{2.5ex} $(2.93\pm 0.13)\times 10^{-6}$ & 1 &  \\
 \cline{2-5}
 & 0.999853 & 21 & \rule{0pt}{2.5ex} $(6.87\pm 0.36)\times 10^{-6}$ & 1 &  \\
 \hline 
\end{tabular}
}
\caption{Intrinsic phase retardation of high reflectivity mirrors. $R$ is the reported reflectivity, $\mathcal{F}$ is the corresponding finesse, $\delta$ their retardation, $N_m$ the number of mirrors of the measured set, and $\lambda$ the wavelength used. The symbol $^\dagger$ indicates data already reported in~\cite{Bielsa2009}.}
\label{tab:BirMir}
\end{table}

\begin{figure}
\centering
\includegraphics[width=\linewidth]{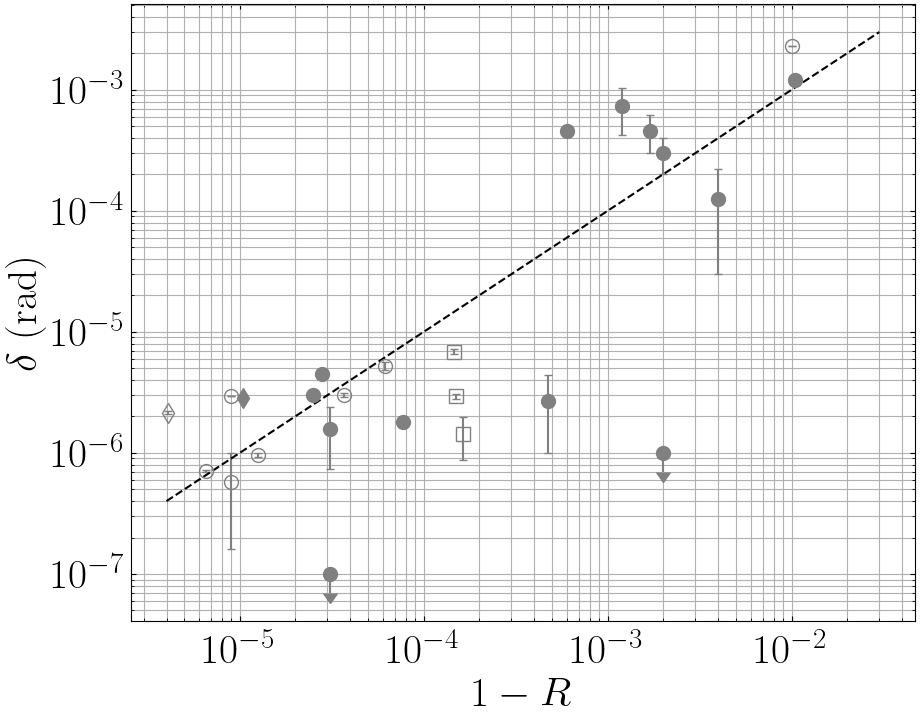}
\caption{Phase retardation per reflection $\delta$ as a function of $1-R$. Filled grey points are the data reported in~\cite{Bielsa2009}, points with arrow represent mirrors with an anisotropy lower than the sensitivity of the apparatus. Empty grey points are data taken from references~\cite{Cadene2015,Lee2000,Morville2002,Huang2008,Yoshino1979,Durand2009,Rivere2017}.  Empty and filled diamonds are data from respectively~\cite{Ejlli2020} et \cite{Kamioka2020}. The three empty square points are taken from reference~\cite{Xiao2019} and were obtained with mirrors of increasing number of deposited layer, the birefringence increasing with the number of layer. To guide the eyes, the trend showed by~\cite{Bielsa2009} is represented by a dashed line}
\label{fig:BruitOpt:Bielsa}
\end{figure}

In Fig.~\ref{fig:BruitOpt:Bielsa} the data of table~\ref{tab:BirMir} are shown as a function of $1-R$ as it was represented in~\cite{Bielsa2009}. All data are given with a coverage factor $k=1$~\cite{JCGM2008}. This kind of plot is however misleading because it shows data as a function of the measured value of $1-R$. In a real mirrors losses are always present, and $1-R=T+P$ with $T$ the transmission of the mirrors and $P$ their losses but Eq.~\ref{eq:BruitOpt:R} assume ideal mirrors, where $P=0$. The simulation producing the B-et-al trend also assumes ideal and lossless mirrors. Thus, comparing the simulated $1-R$ with the experimental one is misleading. We can nevertheless still compare experiments and simulations with the same number $N$ of bilayers. This information is not usually reported explaining why $1-R$ was used in previous studies. Actually, to our knowledge, the number of deposited bilayers $N$ is stated only in~\cite{Xiao2019}. In this work however, the reported reflectivities do not follow Eq.~\ref{eq:BruitOpt:R}. Indeed, assuming $n_s=1.45$ and $n_0=1$, and with $n_H=2.12$ and $n_L=1.48$~\cite{Xiao2019}, we obtain $R_{th}=0.99987$, 0.99997, and 0.9999992 for respectively $N=15$, 17, and 22 the reported number of bilayers. Comparing these values with the one in table~\ref{tab:BirMir}, we see that these mirrors are dominated by losses as the reflectivity does not increase much when adding layers. Actually, we can estimate them as $P=R_{\mathrm{th}}-R$ which gives respectively about $P=38$, 120, and 146~ppm for $N=15$, 17, and 22.

\section{Experimental methods}\label{sec:methods}

Most of the measurements in table~\ref{tab:BirMir} and the new ones we report in this paper have been realised following a similar experimental setup that we recall in the following.

\begin{figure}
\centering
\includegraphics[width=\linewidth]{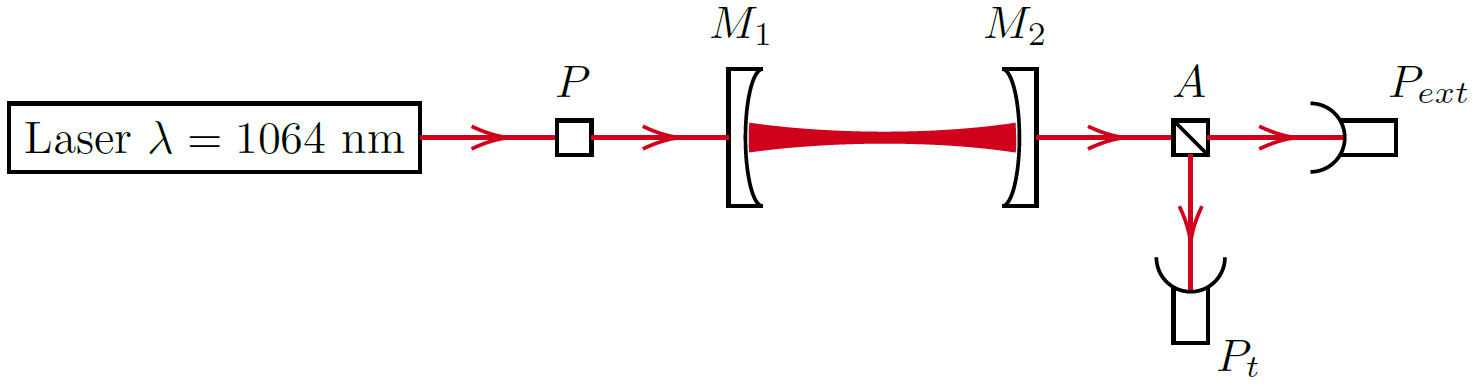}
\caption{Diagram of the polarimeter used to measure mirror birefringence}
\label{fig:BruitOpt:schema}
\end{figure}

In Fig.~\ref{fig:BruitOpt:schema} we show a scheme of the experimental apparatus. It is composed of a laser source, $\lambda=1064$~nm in the case of the new data reported in this work, polarised by a first polariser $P$. This light is incident on a Fabry-Perot cavity composed by two mirrors $M_1$ and $M_2$. The laser is frequency-locked to the TEM$_{00}$ mode of the cavity by means of a Pound-Drever-Hall technique~\cite{Drever1983}. The outgoing light polarisation state is analysed by a second polariser, $A$, crossed with respect to the first one. It is directing the ordinary and extraordinary beam toward two photodiodes, $P_t$ and $P_{ext}$, monitoring the corresponding power levels. With those intensities, we are able to retrieve the ellipticity $\Gamma$ induced by the cavity because of the mirrors intrinsic phase retardation, since
\begin{equation}
\frac{P_{ext}}{P_t}=\sigma^2+\Gamma^2,
\end{equation}
with $\sigma^2$ the extinction ratio of the polarisers accounting for theirs imperfections.

\begin{figure}
\centering
\includegraphics[width=\linewidth]{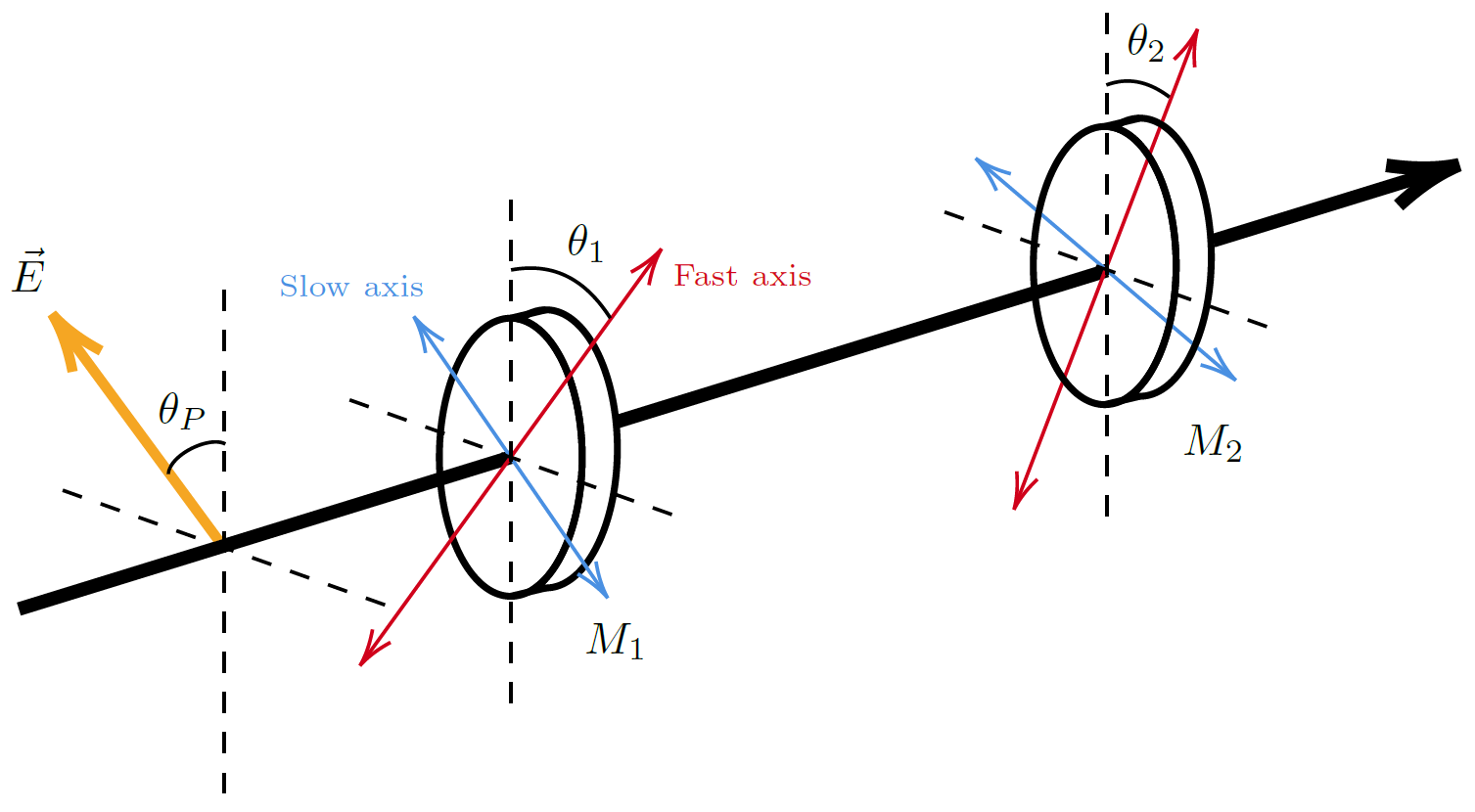}
\caption{Schematic representation of a birefringent cavity. Each mirror has a phase shift between its fast and slow axes, they are rotated by an angle $\theta_i$ relative to a reference. In function of the angle of the incident polarisation $\theta_P$, the acquired ellipticity is equal to Eq.~\ref{eq:etat:Gamma}}
\label{fig:etat:BirCavite}
\end{figure}

To measure the phase retardation of the cavity mirrors, one measures the ellipticity of the cavity as a function of the angle of rotation of each mirror (see Fig.~\ref{fig:etat:BirCavite}). Each mirror is therefore placed on a rotating stage. Indeed, the ellipticity $\Gamma$ of the cavity depends on the angular position $\theta_i$ of each mirror, of their phase retardation $\delta_i$ and of the finesse $\mathcal{F}$ of the cavity. 

To ensure a correct measurement of the cavity finesse, a primary vacuum around the mbar is reached to suppress losses due to O$_2$ absorption~\cite{Bielsa2005}. The measured finesse $\mathcal{F}$ is thus the one of the pair of mirrors and not the one of each individual mirror. The cavity finesse can be written as
\begin{equation}
\frac{\mathcal{F}}{\pi}=\frac{1}{1-r_1 r_2}
\end{equation}
where $r_i$ is the coefficient of reflection in amplitude of the mirror $i$. If we define the finesse $\mathcal{F}_i$ of each mirror as
\begin{equation}
\frac{\mathcal{F}_i}{\pi}=\frac{1}{1-r_i^2},
\end{equation}
then the cavity finesse is
\begin{equation}\label{eq:BruitOpt:Fasym}
\frac{\mathcal{F}}{\pi}=\frac{1}{1-\sqrt{\left(1-\dfrac{\pi}{\mathcal{F}_1}\right)\left(1-\dfrac{\pi}{\mathcal{F}_2}\right)}}.
\end{equation}
The cavity finesse is usually measured thanks to the photon lifetime in the cavity $\tau$ which is the characteristic time of the exponential decay observed by the transmitted power when the incident light of the cavity is suddenly turned off. It is equal to $\mathcal{F}=\pi c\tau/L_c$ with $c$ the light celerity and $L_c$ the length of the cavity.

\begin{figure}
\centering
\includegraphics[width=\linewidth]{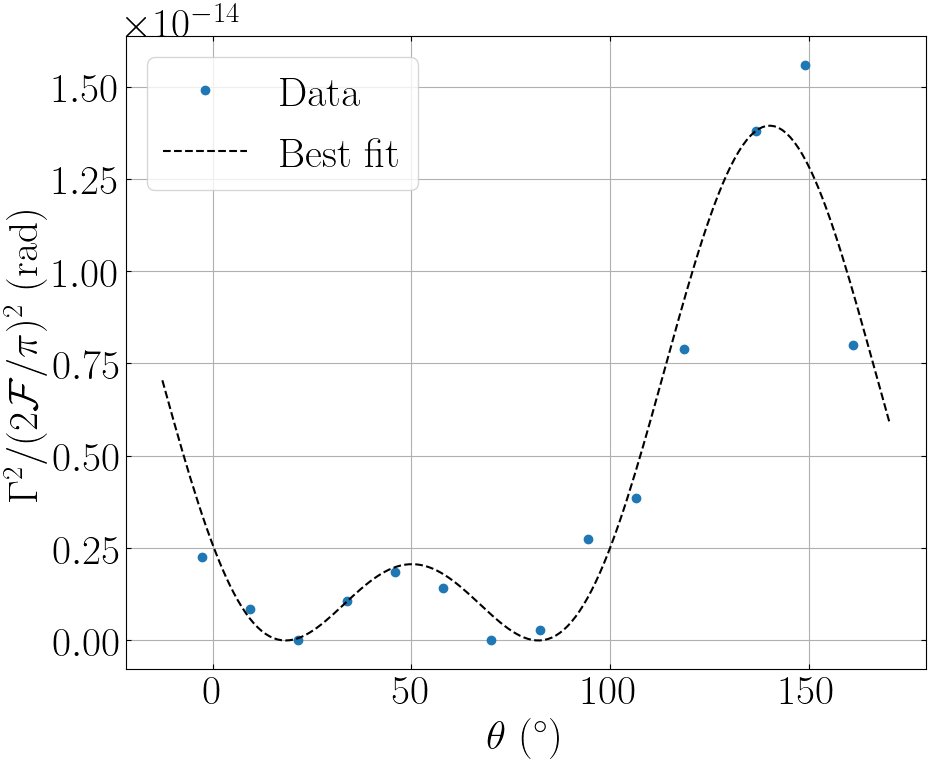}
\caption{Typical experimental points of one mirror used in this work and its best fit with Eq.~\ref{eq:BruitOpt:fit}}
\label{fig:BruitOpt:Fit}
\end{figure}

The principle of the measurement is the following. A mirror is turned by an angle $\theta$ and we note the power $P_t$ and $P_{ext}$ as well as the photon lifetime $\tau$. With these quantities we can compute the quantity $\Gamma^2/(2\mathcal{F}/\pi)^2$ in function of $\theta$. Since the ellipticity $\Gamma$ can be expressed as
\begin{equation}\label{eq:etat:Gamma}
\Gamma=\frac{2\mathcal{F}}{\pi}\left(\frac{\delta_1}{2}\sin(2(\theta_1-\theta_P))+\frac{\delta_2}{2}\sin(2(\theta_2-\theta_P))\right),
\end{equation}
thanks to a least square fit by a function of the form
\begin{equation}\label{eq:BruitOpt:fit}
f(\theta)=\left(A\sin (2(\theta-\theta_0)) +B\right)^2
\end{equation}
we can retrieve $\delta=2A$ and the angle $\theta_0$ between the optical axis of the mirror and the incident polarisation. An example of data thus obtained and their best fit is shown in Fig.~\ref{fig:BruitOpt:Fit}. This method is not sensitive to the sign of $\delta$, and it will be considered positive in the following. Let us note that flipping the sign of $\delta$ is equivalent to adding $\pi/2$ to the angle $\theta_0$.

\section{Measurements}
\subsection{FiveNine Optics mirrors}

Using the method detailed in section~\ref{sec:methods}, we have characterised some commercial mirrors of radius of curvature $C=2$~m, made by FiveNine Optics and used by the BMV experiment in~\cite{Beard2021}, with a cavity of length $L_c=2.55$~m. They are reported in table~\ref{tab:BruitOpt:resultatFNO} as well as in Fig.~\ref{fig:BruitOpt:Bielsa2}. A second measurement of the same mirrors at a lower finesse is also shown in table~\ref{tab:BruitOpt:resultatFNO}. The cause of this loss of finesse is a pollution after several cycling between vacuum and atmospheric pressure. 

\begin{table}
\centering
\begin{tabular}{|c|c|c|c|}
 \hline
 \rule{0pt}{2.5ex} Mirror  & $R$ & $\mathcal{F}\times 10^{-3}$ & $\delta\times 10^{7}$ (rad) \\
 \hline
 \hline
 \multirow{2}{*}{FNO}  & \rule{0pt}{2.5ex}$0.9999889(2)$ & $283\pm 5$ & $6.5\pm 0.4$  \\
 \cline{2-4}
    & \rule{0pt}{2.5ex}$0.9999869(2)$ & $240\pm 4$ & $5.2\pm 0.9$  \\
 \hline
 \hline
 \multirow{2}{*}{FNO$^\dagger$}  & \rule{0pt}{2.5ex}$0.999818(3)$ & $17.3\pm 0.3$ & $32.6\pm 0.8$  \\
 \cline{2-4}
   & \rule{0pt}{2.5ex}$0.999818(3)$ & $17.3\pm 0.3$ & $7.2\pm 0.3$ \\
 \hline
\end{tabular}
\caption{Results for each mirrors during the first round of measurement. $R$ is the mean reflectivity, $\mathcal{F}$ is the mean finesse, and $\delta$ is the phase retardation per reflection. Uncertainties are given with a coverage factor $k=1$. FNO designate a pair of commercial mirror, FNO$^\dagger$ are the same mirrors with a lower finesse, see text}
\label{tab:BruitOpt:resultatFNO}
\end{table}

Notwithstanding the fact that $1-R$ has increased by the pollution, the variation of the mirror phase retardation did not follow the one expected with a lower finesse by the B-et-al trend. This shows the fact that by adding superficial losses, one does not change the intrinsic phase retardation of the mirrors. The number of layers of the mirrors being unchanged, this confirms the fact that $\delta$ depends on $N$ and not on the measured $1-R$.
 
\subsection{Safran Electronics \& Defense mirrors}

To understand the origin of this mirror phase retardation, we have performed a larger-scale study to investigate it, measuring several mirrors done for this purpose. Since the hypothesis of~\cite{Bielsa2009} is that the birefringence is contained in the first layer in contact with the substrate, our study looked at the influence of the substrate by depositing on different substrates. Another goal was to vary the number of layers deposited on the same mirrors. Therefore, our study was carried out in two stages. First, an initial series of measurements was carried out on the mirrors. Then, additional layers were deposited on the same mirrors for a second series of measurements.

\subsubsection{First round of measurement}

To test the substrate influence, we used silica substrates from different companies, Coastline, Edmund Optics and Laseroptik, and some Zerodur substrates. All the substrates except the Coastline ones were polished by Safran, and they also carried out the thin film deposition on all the substrates at the same time in the same conditions. During the deposition, the substrates were all aligned according to a special mark etched on the mirror substrates. During the installation of a cavity for measurement, we placed the mirrors on their rotating stage using this mark as a reference point. We calibrated the angle of the rotating stage with a waveplate of known optical axis that we aligned along the same reference point. This allows us to know the angle between the reference point and the incident polarisation, and consequently between the mark of a substrate and the incident polarisation. This was done for both rotating stage of our Fabry-Perot cavity whose length is $L_c=1.23$~m.

We have performed the measurement on 10 pairs of mirrors, whose radius of curvature was $C=1$~m: 4 Coastline, 6 Zerodur, 5 Edmund Optics and 5 Laseroptik as far as the substrate is concerned. They are respectively designated in the following by the prefixes ``CO'', ``B0'', ``ED'', and ``LK''. Because we calibrated the rotating angle of the mirror with a common reference point, we can compare the angle of the optical axis $\theta_0$ between mirrors. The measured mean reflectivity $R$, phase retardation $\delta$ and optical axis angle $\theta_0$ are grouped in table~\ref{tab:BruitOpt:resultat} and sorted two by two according to their assignation to form a Fabry-Perot cavity. In the table, the uncertainties on the parameters $\delta$ and $\theta_0$ come from the fitting procedure, the one of $R$ come from the dispersion of the photon lifetime values collected during the rotation of the mirror.

\begin{table}
\centering
\resizebox{\linewidth}{!}{
\begin{tabular}{|c|c|c|c|c|}
 \hline
 \rule{0pt}{2.5ex} Mirror  & $R$ & $\mathcal{F}\times 10^{-3}$ & $\delta\times 10^{7}$ (rad) & $\theta_0$ ($^\circ$)\\
 \hline
 \hline
 \rule{0pt}{2.5ex} CO1624  & $0.999899(3)$ & $31.4\pm 0.8$ & $8.8\pm 0.2$ & $40.2\pm 1.2$ \\
 \hline
 \rule{0pt}{2.5ex} CO1625  & $0.999893(6)$ & $30.8\pm 1.6$ & $1.3\pm 0.5$ & $-5.0\pm 12.0$ \\
 \hline
 \hline
 \rule{0pt}{2.5ex} CO1629  & $0.999859(2)$ & $22.3\pm 0.4$ & $4.6\pm 0.3$ & $6.2\pm 1.4$ \\
 \hline
 \rule{0pt}{2.5ex} CO1598 & $0.999849(2)$ & $20.8\pm 0.3$ & $10.7\pm 0.6$ & $39.9\pm 1.5$ \\
 \hline
 \hline
 \rule{0pt}{2.5ex} B0-94063  & $0.9999897(2)$ & $307\pm 5$ & $0.49\pm 0.06$ & $14.3\pm 2.3$ \\
 \hline
  \rule{0pt}{2.5ex} B0-94265  & $0.9999881(7)$ & $271\pm 14$ & $2.4\pm 0.1$ & $15.5\pm 1.9$ \\
 \hline
 \hline
  \rule{0pt}{2.5ex} B0-94056 & $0.9999905(1)$ & $333\pm 5$ & $1.4\pm 0.2$ & $16.4\pm 3.3$ \\
 \hline
   \rule{0pt}{2.5ex} B0-94013  & $0.99999043(8)$ & $329\pm 3$ & $0.9\pm 0.1$ & $1.8\pm 2.0$ \\
 \hline
 \hline
   \rule{0pt}{2.5ex} B0-94060  & $0.9999904(1)$ & $327\pm 4$ & $2.5\pm 0.2$ & $4.8\pm 2.5$ \\
 \hline
   \rule{0pt}{2.5ex} B0-94034  & $0.99999029(6)$ & $324\pm 2$ & $0.97\pm 0.17$ & $5.0\pm 4.0$ \\
 \hline
 \hline
   \rule{0pt}{2.5ex} ED08  & $0.9999920(1)$ & $392\pm 5$ & $1.6\pm 0.3$ & $21.0\pm 4.0$ \\
 \hline
    \rule{0pt}{2.5ex} ED12  & $0.99999205(9)$ & $396\pm 5$ & $1.22\pm 0.09$ & $29.1\pm 1.2$ \\
 \hline
 \hline
    \rule{0pt}{2.5ex} ED06  & $0.99999218(9)$ & $403\pm 4$ & $1.4\pm 0.2$ & $26.0\pm 3.3$ \\
 \hline
    \rule{0pt}{2.5ex} ED07  & $0.9999914(1)$ & $368\pm 5$ & $1.9\pm 0.3$ & $24.9\pm 2.7$ \\
 \hline
 \hline
    \rule{0pt}{2.5ex} ED05  & $0.99999298(8)$ & $448\pm 5$ & $1.0\pm 0.1$ & $12.0\pm 4.0$ \\
 \hline
    \rule{0pt}{2.5ex} LK04  & $0.99999262(9)$ & $427\pm 5$ & $1.64\pm 0.09$ & $5.2\pm 1.5$ \\
 \hline
 \hline
 \rule{0pt}{2.5ex} LK10  & $0.99999241(7)$ & $414\pm 4$ & $3.2\pm 0.2$ & $19.6\pm 2.9$ \\
 \hline
 \rule{0pt}{2.5ex} LK11  & $0.9999923(1)$ & $411\pm 6$ & $1.8\pm 0.2$ & $-13.3\pm 2.2$ \\
 \hline
 \hline
 \rule{0pt}{2.5ex} LK01  & $0.9999920(1)$ & $393\pm 5$ & $2.3\pm 0.2$ & $34.5\pm 2.9$ \\
 \hline
 \rule{0pt}{2.5ex} LK05  & $0.99999077(7)$ & $341\pm 3$ & $1.6\pm 0.3$ & $13.0\pm 5.0$ \\
 \hline
\end{tabular}
}
\caption{Results for each mirrors during the first round of measurement. $R$ is the mean reflectivity, $\mathcal{F}$ is the mean finesse, $\delta$ is the phase retardation, and $\theta_0$ is the direction of the optical axis with respect to the mark on the substrate. Uncertainties are given with a coverage factor $k=1$. Prefix CO, ED and LK refer respectively to the Coastline, Edmund Optics and Laseroptik substrates. Prefix B0 refer to Zerodur substrates. Mirror LK04 is the one whose experiment points have been shown in Fig.~\ref{fig:BruitOpt:Fit}}
\label{tab:BruitOpt:resultat}
\end{table}

\begin{figure}
\centering
\includegraphics[width=\linewidth]{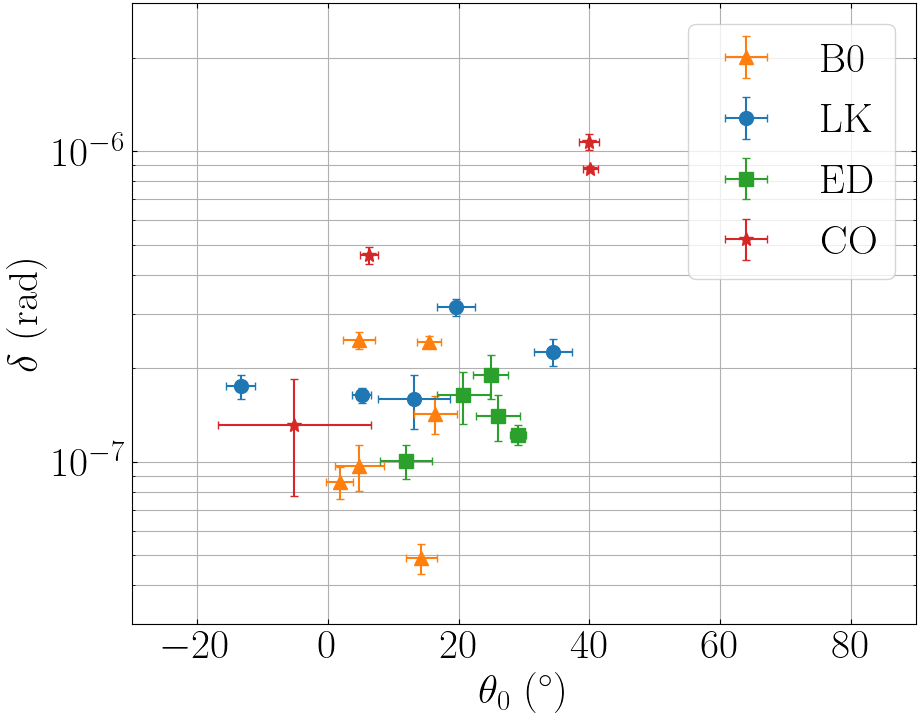}
\caption{Phase retardation $\delta$ as a function of the direction of the optical axis $\theta_0$ with respect to the mark of each mirror. Uncertainties are given with a coverage factor $k=1$}
\label{fig:BruitOpt:DeltaTheta}
\end{figure}

In Fig.~\ref{fig:BruitOpt:DeltaTheta} we represent for each mirror the value of the phase retardation as a function of the direction of its optical axis. We observe that the direction of the optical axes gathering around the value $\theta_0\sim 10^\circ$ indicating a preferred direction during the mirror manufacturing process.

There are differences in the amplitude of birefringence depending on the substrate. Coastline is generally higher than Laseroptik, Edmund and Zerodur. This seems to validate the role of the substrate on the phase retardation of the mirror, in particular the role of the polishing stage.

\begin{figure}
\centering
\includegraphics[width=\linewidth]{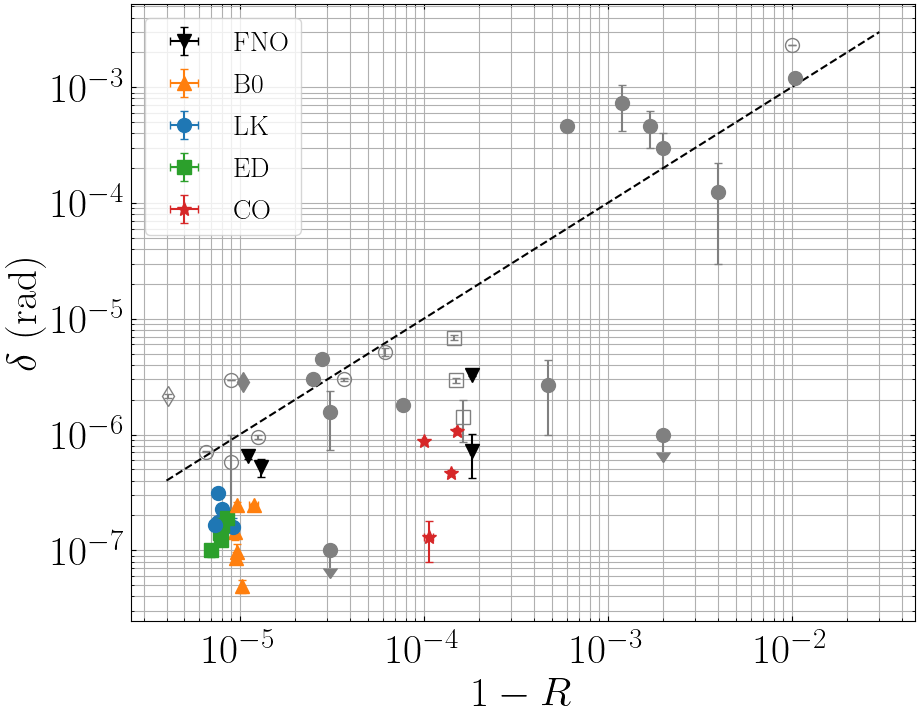}
\caption{Phase retardation per reflection $\delta$ as a function of $1-R$. Colored points are the new measurements reported in this article. Other points are those previously shown in Fig.~\ref{fig:BruitOpt:Bielsa}}
\label{fig:BruitOpt:Bielsa2}
\end{figure}

Figure~\ref{fig:BruitOpt:Bielsa2} shows the phase retardation $\delta$ as a function of $1-R$ of the mirrors that we have measured as well as the B-et-al trend. Again, this kind of plot can be misleading. As a matter of fact, although
all mirrors were deposited in the same conditions, in particular with the same number of layers, the Coastline mirrors present a smaller reflectivity. This is explained by the fact that these substrates presented worse surface quality before deposition. If one would shift these points towards the others mirrors, one would see that their phase retardation is for 3 of them in accordance with the B-et-al trend.

We observe that the phase retardation of the other mirrors, polished by Safran, are grouped around a few $10^{-7}$~rad, lower than the trend of~\cite{Bielsa2009}. However, the same uncertainties on the position of the abscissa of the points apply because, as we previously argued, the B-et-al trend does not account for the losses $P$.

\subsubsection{Second round of measurement}

After the first round of measurement, we deposited two additional pairs of layers on the same mirrors.

Once the new deposition done, we reiterated the same measurements as before using the same methodology, in particular the same pairings. The new results are reported in table~\ref{tab:BruitOpt:resultatNew}.

\begin{table}
\centering
\resizebox{\linewidth}{!}{
\begin{tabular}{|c|c|c|c|c|}
 \hline
 \rule{0pt}{2.5ex} Mirror & $R$ & $\mathcal{F}\times 10^{-3}$ & $\delta\times 10^7$ (rad) & $\theta_0$ ($^\circ$)\\
 \hline
 \hline
 \rule{0pt}{2.5ex} B0-94063  & $0.9999944(2)$ & $568\pm 13$ & $0.92\pm 0.06$ & $-9.0\pm 3.0$ \\
 \hline
  \rule{0pt}{2.5ex} B0-94265  & $0.99999449(9)$ & $573\pm 9$ & $3.5\pm 0.2$ & $11.6\pm 3.0$ \\
 \hline
 \hline
  \rule{0pt}{2.5ex} B0-94056  & $0.99999468(9)$ & $593\pm 9$ & $2.0\pm 0.5$ & $23.6\pm 3.4$ \\
 \hline
   \rule{0pt}{2.5ex} B0-94013  & $0.9999938(5)$ & $528\pm 26$ & $1.4\pm 0.1$ & $25.0\pm 4.0$ \\
 \hline
 \hline
   \rule{0pt}{2.5ex} ED08  & $0.9999950(1)$ & $627\pm 12$ & $1.02\pm 0.07$ & $1.0\pm 3.1$ \\
 \hline
    \rule{0pt}{2.5ex} ED12  & $0.9999948(1)$ & $608\pm 15$ & $0.8\pm 0.1$ & $8.0\pm 4.0$ \\
 \hline
 \hline
    \rule{0pt}{2.5ex} ED06  & $0.99999601(6)$ & $789\pm 11$ & $2.9\pm 0.2$ & $24.7\pm 2.1$ \\
 \hline
    \rule{0pt}{2.5ex} ED07  & $0.9999929(2)$ & $449\pm 11$ & $3.8\pm 0.2$ & $17.6\pm 1.3$ \\
 \hline
 \hline
 \rule{0pt}{2.5ex} LK10  & $0.9999952(2)$ & $676\pm 29$ & $2.4\pm 0.3$ & $24.0\pm 4.0$ \\
 \hline
 \rule{0pt}{2.5ex} LK11  & $0.9999943(3)$ & $570\pm 24$ & $1.3\pm 0.1$ & $17.0\pm 4.0$ \\
 \hline
 \hline
 \rule{0pt}{2.5ex} LK01  & $0.999966(2)$ & $101\pm 7$ & $7.8\pm 0.4$ & $22.1\pm 1.9$ \\
 \hline
 \rule{0pt}{2.5ex} LK05  & $0.9999945(2)$ & $587\pm 24$ & $1.7\pm 0.3$ & $-2.0\pm 4.0$ \\
 \hline
\end{tabular}
}
\caption{Results for each mirror with two additional pairs of layers. $R$ is the mean reflectivity, $\mathcal{F}$ is the mean finesse, $\delta$ is the phase retardation per reflection and $\theta_0$ is the direction of the optical axis with respect to the mark on the substrate. Uncertainties are given with a coverage factor $k=1$. Prefix CO, ED and LK refer respectively to the Coastline, Edmund Optics and Laseroptik substrates. Prefix B0 refer to Zerodur substrates}
\label{tab:BruitOpt:resultatNew}
\end{table}

With the additional layers, we measured, as expected, an increase in the reflectivity $R$ of the mirrors. In fact, in terms of finesse, we measured a maximum finesse of 895~000. To our knowledge, this constitute a record for our operating wavelength, $\lambda=1064$~nm, and the second highest reported finesse overall, as one can see in table~\ref{tab:BruitOpt:finesse} which regroup the performances of various Fabry-Perot cavities.

\begin{table}
\centering
\begin{tabular}{|c|c|c|c|}
 \hline
 \rule{0pt}{2.5ex} Cavity & $\lambda$ (nm) & $L_c$ (m) & $\mathcal{F}$ \\
 \hline
 \rule{0pt}{2.5ex} ALPS II~\cite{Pold2020} & 1064 & 9.2 & 101~300\\
 \hline
 \rule{0pt}{2.5ex} BMV~\cite{Beard2021} & 1064 & 2.55 & 537~000\\
 \hline
 \rule{0pt}{2.5ex} OVAL~\cite{Fan2017} & 1064 & 1.38 & 670~000\\
 \hline
 \rule{0pt}{2.5ex} PVLAS~\cite{DellaValle2014} & 1064 & 3.303 & 770~000\\
 \hline
 \rule{0pt}{2.5ex} J. Millo~\emph{et~al.}~\cite{Millo2009} & 1064 & 0.1 & 800~000\\
 \hline
 \rule{0pt}{2.5ex} This work & 1064 & 1.23 & 895~000\\
 \hline
 \rule{0pt}{2.5ex} G. Rempe~\emph{et~al.}~\cite{Rempe1992} & 850 & 0.004 & 1~900~000\\
 \hline
\end{tabular}
\caption{Wavelength, cavity length, and Fabry-Perot cavity finesse of high finesse cavities}
\label{tab:BruitOpt:finesse}
\end{table}

Considering the pair giving the record finesse, the average finesse for the first deposition of 19 pairs of layers was $\mathcal{F}=413~000\pm 3~500$, corresponding to a coefficient of reflection $1-R=(7.62\pm 0.07)\times 10^{-6}$. If we know the power entering and leaving the cavity, we can determine its transmittance, $T_c=0.114\pm 0.04$. With the latter, if the mirrors are identical, \emph{i.e.} $r=r_1=r_2$ and $t=t_1=t_2$, we have 
\begin{equation}\label{eq:etat:TcRc}
T_c=\left(T\frac{\mathcal{F}}{\pi}\right)^2 \quad \text{and} \quad R_c=R\left(1-T\frac{\mathcal{F}}{\pi}\right)^2,
\end{equation}
with $T=t^2$ and $R=r^2$. We can thus deduce the value of $T$, $T=(2.57\pm 0.05)\times 10^{-6}$. Finally, because $R+T+P=1$ we can estimate the losses of the pair as $P=(5.06\pm 0.08)\times 10^{-6}$.

We did the same analysis after the deposition of two additional pairs of layers, we obtained a mean finesse of $\mathcal{F}=623~000\pm 19~000$ corresponding to $1-R=(5.2\pm 0.2)\times 10^{-6}$. The cavity transmittance was $T_c=0.028\pm 0.002$, we thus had $T=(8.7\pm 0.3)\times 10^{-7}$, and $P=(4.4\pm 0.2)\times 10^{-6}$. As one expect, with the additional layers, $1-R$ and $T$ decreased, interestingly the losses $P$ also slightly decreased.

\begin{figure}
\centering
\includegraphics[width=\linewidth]{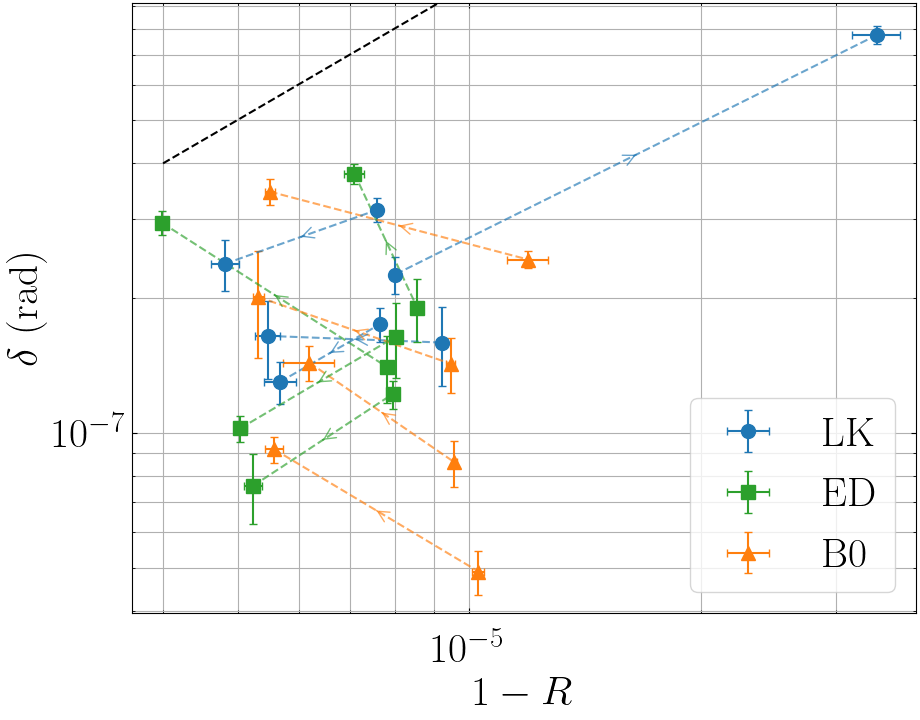}
\caption{Phase retardation per reflection $\delta$ as a function of $1-R$. Measurements before and after the deposition of two additional pairs of layers are linked by an arrow. Data from mirrors that were not measured twice are not reported}
\label{fig:BruitOpt:Bielsa3}
\end{figure}

From the point of view of the phase retardation, we observe in Fig.~\ref{fig:BruitOpt:Bielsa3} that the latter does not show a unique behaviour. Indeed, in our sample of twelve mirrors, four of them have had their birefringence reduced, one stayed constant and the rest increased. Following the B.et.at trend, we expected only a diminution of the phase retardation.

\begin{figure}
\centering
\includegraphics[width=\linewidth]{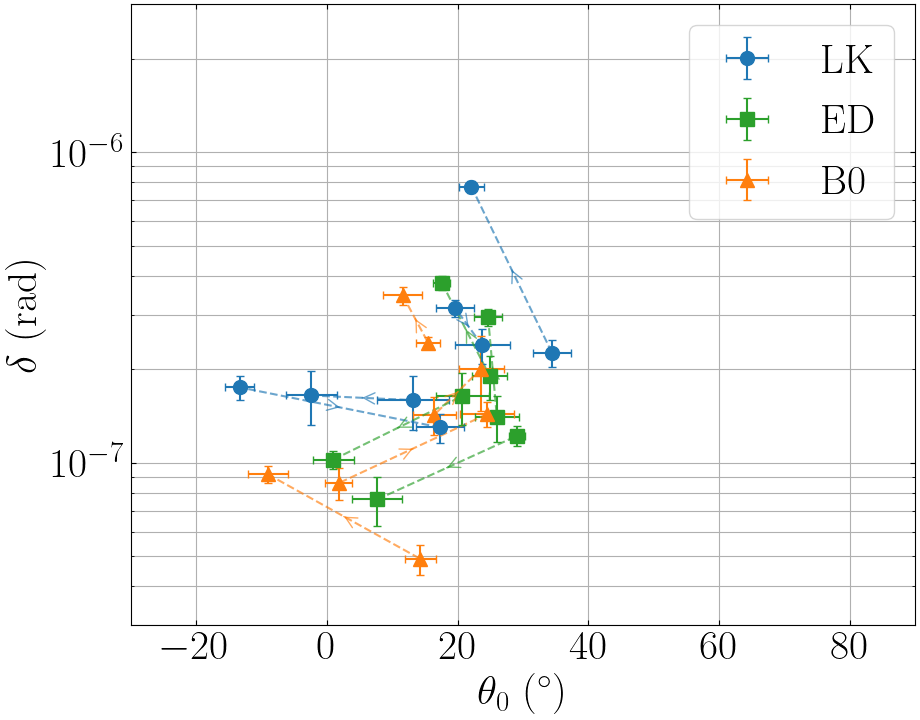}
\caption{Phase retardation per reflection $\delta$ as a function of the direction of the optical axis $\theta_0$ with respect to the mark of each substrate. Measurements before and after the deposition of two additional pairs of layers are linked by an arrow}
\label{fig:BruitOpt:deltathetaNew}
\end{figure}

In addition, we note in Fig.~\ref{fig:BruitOpt:deltathetaNew} that all the optical axes were displaced by adding the layers. This fact seems to indicate that by their addition, we also added a waveplate. Indeed, if we consider two waveplates of retardation $\delta_i$ and optical axes making an angle $\theta_i$ with the incident polarisation, then the waveplate equivalent to the combination of the two is characterised by a $\delta_{eq}$ and $\theta_{eq}$ such that~\cite{Brandi1997} 
\begin{equation}\label{eq:bruitopt:lameeq}
\begin{split}
&\delta_{eq}=\sqrt{\left(\delta_1-\delta_2\right)^2+4\delta_1\delta_2\cos^2(\theta_2-\theta_1)}\\
&\cos(2(\theta_{eq}-\theta_1))=\frac{\delta_1+\delta_2\cos(2(\theta_2-\theta_1))}{\delta_{eq}}.
\end{split}
\end{equation}
In our case, we performed a first round of measurements determining $\delta_1$ and $\theta_1$ of Eq.~\ref{eq:bruitopt:lameeq}. Then, we carried out a second round of measurements with the added layers, determining the birefringence of the combination of the old measurements and the added layers, \emph{i.e.} $\delta_{eq}$ and $\theta_{eq}$. We can thus determine the phase retardation of only the added layers by inverting Eq.~\ref{eq:bruitopt:lameeq}. We obtain
\begin{equation}\label{eq:bruitopt:lame2}
\begin{split}
&\delta_2=\pm\sqrt{\delta_1^2+\delta_{eq}^2-2\delta_1\delta_{eq}\cos(2(\theta_{eq}-\theta_1))}\\
&\cos(2(\theta_2-\theta_1))=\frac{\delta_{eq}\cos(2(\theta_{eq}-\theta_1))-\delta_1}{\delta_2}.
\end{split}
\end{equation}
Let us note that the two solutions $\delta_2$ of opposite signs present a $\theta_2-\theta_1$ separated by $\pi/2$, they thus represent the same waveplate. Indeed, changing the sign of a waveplate phase retardation, means swapping the fast and slow axes, \emph{i.e.} turning it by $\pi/2$.

\begin{figure}
\centering
\includegraphics[width=\linewidth]{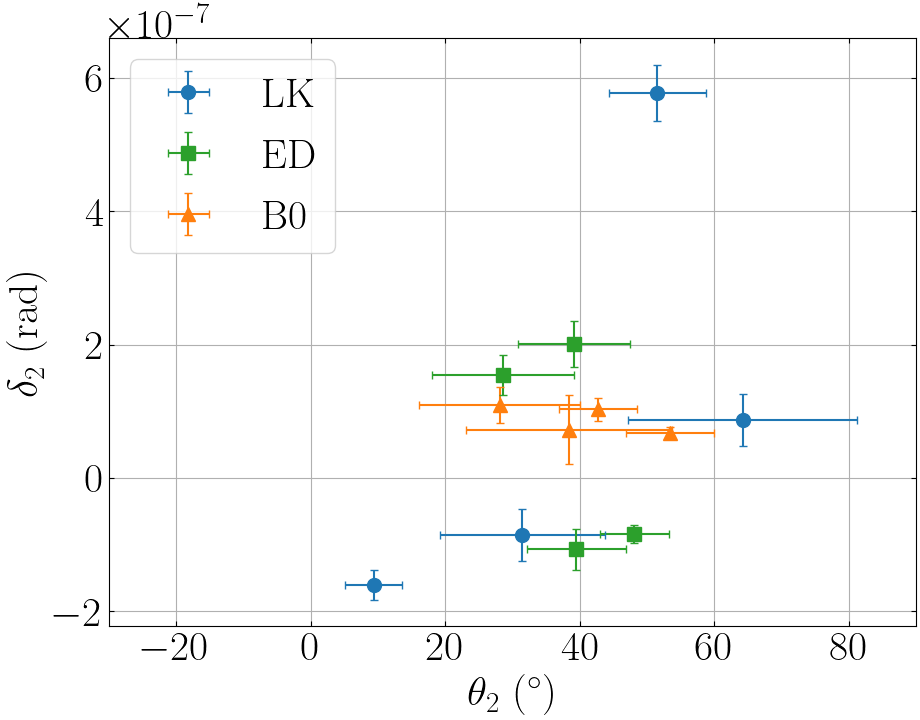}
\caption{Phase retardation per reflection of the equivalent retardation plate of the additional layers $\delta_2$ as a function of their direction of the optical axis $\theta_2$ with respect to the mark on each substrate}
\label{fig:BruitOpt:delta2theta2}
\end{figure}

In Fig.~\ref{fig:BruitOpt:delta2theta2}  we show the values thus obtained of $\delta_2$ and $\theta_2$ with our data. To ensure that the angles $\theta_2$ obtained are in the same range as those of the Fig.~\ref{fig:BruitOpt:DeltaTheta} and \ref{fig:BruitOpt:deltathetaNew} we had to change the sign of the phase retardation of some mirrors. It is interesting to notice that this happened for mirrors whose birefringence decreased, \emph{i.e.} those where $\delta_1>\delta_{eq}$. In absolute value, the phase retardation of all the added layers are $\delta_2=(1.5\pm 0.4)\times 10^{-7}$~rad.

To reproduce the shape of the trend in $\delta\varpropto 1-R$, in the simulation of the article~\cite{Bielsa2009} only the first layer in contact with the substrate is birefringent. Nevertheless, our measurement seems to indicate that one has to attribute a phase retardation of the order of $\delta\lesssim 10^{-7}$~rad to each of them. This contribution should be added to the model. One can expect that for a large number of layers, where the impact of the first layer is null, only the last layers contribute to the global phase retardation. Thus, adding a layer brings its intrinsic phase retardation but remove the same amount in vanishing the electric fields earlier in the multilayer. One can thus expect a constant phase retardation when adding layers if their optical axes are aligned, as it was showed in our study.


\section{Computational study}
\subsection{Methods}
We want to extend the computational study of~\cite{Bielsa2009} to take into account the birefringent nature of each layer. To do so we use the same characteristic matrix method~\cite{Born1983} that was introduced in~\cite{Bielsa2009}. In this formulation we can describe the multilayer by a matrix $M$ relating the amplitude of the outgoing and ingoing electric field both before and after the multilayer. Thus, $M$ is a $4\times 4$ matrix such as~\cite{Bielsa2009}
\begin{equation}
\begin{pmatrix}
A_{e,x}^+\\
A_{e,x}^-\\
A_{e,y}^+\\
A_{e,y}^-
\end{pmatrix}=M\begin{pmatrix}
A_{s,x}^+\\
A_{s,x}^-\\
A_{s,y}^+\\
A_{s,y}^-
\end{pmatrix},
\end{equation}
where we have considered that the light electric field propagate through the $z$ axis, $e$ and $s$ refer respectively to the external medium and to the substrate and the sign $\pm$ refers to the direction of propagation of the field. In this notation, $A_{e,x}^+$ and $A_{e,y}^+$ are the two components of the incident field, $A_{e,x}^-$ and $A_{e,y}^-$ are the components of the reflected field, and $A_{s,x}^+$ and $A_{s,y}^+$ are the ones of the transmitted field. Lastly, $A_{s,x}^-$ and $A_{s,y}^-$ are the components of the field going into the stack from the substrate side.

\begin{figure}
\centering
\includegraphics[width=\linewidth]{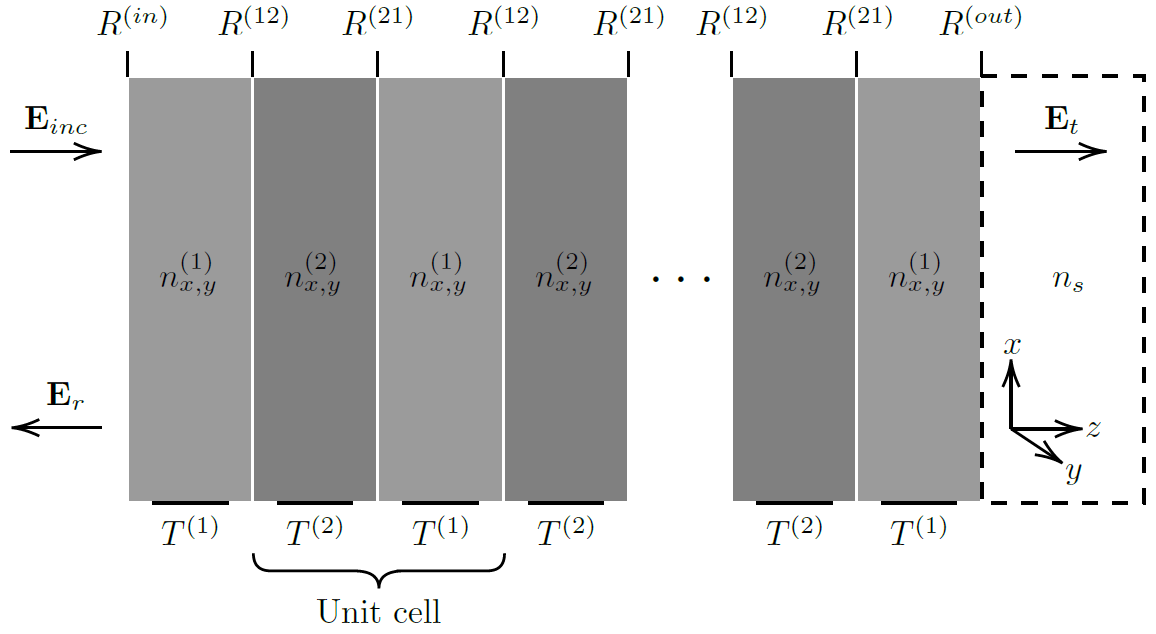}
\caption{Scheme of a multilayer mirror}
\label{fig:Stack}
\end{figure}

By the stratified nature of the multilayer system that we are considering, the matrix $M$ can be decomposed as the product of matrices corresponding to the transfer matrices at the different interfaces and propagation matrices between them. The multilayer is composed of alternating layers of refractive index $n^{(1)}$ and $n^{(2)}$ as shown in Fig.~\ref{fig:Stack}. Propagation through a layer can be described by the matrix $T^{(i)}$
\begin{equation}
T^{(i)}=\begin{pmatrix}
e^{-i n_x^{(i)} kd^{(i)}} & 0 & 0 & 0\\
0 & e^{i n_x^{(i)} kd^{(i)}} & 0 & 0\\
0 & 0 & e^{-i n_y^{(i)} kd^{(i)}} & 0\\
0 & 0 & 0 & e^{i n_y^{(i)} kd^{(i)}}
\end{pmatrix},
\end{equation}
where $n^{(i)}_{x,y}$ is the index of refraction of the material $i$ in the direction $x$ and $y$, $k$ is the wave vector of the electric field and $d^{(i)}$ is the thickness of each layer. It is chosen so that $n^{(1)}d^{(1)}=n^{(2)}d^{(2)}=\lambda/4$ assuring constructive interference and maximum reflectivity. In practice, because we will consider that the layers are birefringent, we assume that this condition is met for the $x$ direction \emph{i.e.} that $kd^{(i)}=\pi/(2n_x^{(i)})$.

Following~\cite{Kryhin2023}, thanks to the boundary conditions between the material 1 and 2, one can show that the transfer matrix of the interface is
\begin{equation}
R^{(12)}=\frac{1}{2}\begin{pmatrix}
1+a_x^{(12)} & 1-a_x^{(12)} & 0 & 0\\
1-a_x^{(12)} & 1+a_x^{(12)} & 0 & 0\\
0 & 0 & 1+a_y^{(12)} & 1-a_y^{(12)}\\
0 & 0 & 1-a_y^{(12)} & 1+a_y^{(12)}
\end{pmatrix},
\end{equation}
with $a_{x,y}^{(ij)}=n^{(j)}_{x,y}/n^{(i)}_{x,y}$. To obtain the transfer matrix $R^{(21)}$, one swap the materials \emph{i.e.} $1\leftrightarrow 2$.

The multilayer is composed of a stack of a bilayer of material 1 and 2, the transfer matrix of this basic element is 
\begin{equation}
L=R^{(12)}T^{(2)}R^{(21)}T^{(1)}.
\end{equation}
Because the coating is composed of $N$ bilayers, the total transfer matrix of the system is
\begin{equation}
M=R^{(in)}T^{(1)}L^NR^{(out)},
\end{equation}
where $R^{(in)}$ and $R^{(out)}$ are the transfer matrices of the entrance interface and the substrate one. Thus,
\begin{equation}
R^{(in)}=\frac{1}{2}\begin{pmatrix}
1+a_x^{(e1)} & 1-a_x^{(e1)} & 0 & 0\\
1-a_x^{(e1)} & 1+a_x^{(e1)} & 0 & 0\\
0 & 0 & 1+a_y^{(e1)} & 1-a_y^{(e1)}\\
0 & 0 & 1-a_y^{(e1)} & 1+a_y^{(e1)}
\end{pmatrix},
\end{equation}
where $a_{x,y}^{(e1)}=n^{(1)}_{x,y}$ considering that the index of refraction of the external medium is equal to 1. We also have
\begin{equation}
R^{(out)}=\frac{1}{2}\begin{pmatrix}
1+a_x^{(1s)} & 1-a_x^{(1s)} & 0 & 0\\
1-a_x^{(1s)} & 1+a_x^{(1s)} & 0 & 0\\
0 & 0 & 1+a_y^{(1s)} & 1-a_y^{(1s)}\\
0 & 0 & 1-a_y^{(1s)} & 1+a_y^{(1s)}
\end{pmatrix},
\end{equation}
where $a_{x,y}^{(1s)}=n_s/n^{(1)}_{x,y}$ with $n_s$ the index of refraction of the substrate. This expression of $R^{(out)}$ is slightly different from the one derived in~\cite{Kryhin2023} where $n_s$ and $n^{(1)}_{x,y}$ where swapped and material 2 was used instead of 1.

In our derivation, we considered that the optical axes of each layer are aligned. Once the matrix $M$ is computed, we can compute the transmitted and reflected amplitudes. Indeed, considering that the incident polarisation is linear and make an angle $\theta$ with the $x$ axis, we have
\begin{equation}
\begin{pmatrix}
E_0\cos\theta\\
E_x^r\\
E_0\sin\theta\\
E_y^r
\end{pmatrix}=\begin{pmatrix}
M_{11} & M_{12} & M_{13} & M_{14} \\ 
M_{21} & M_{22} & M_{23} & M_{24} \\ 
M_{31} & M_{32} & M_{33} & M_{34} \\ 
M_{41} & M_{42} & M_{43} & M_{44}  
\end{pmatrix}\begin{pmatrix}
E_x^t\\
0\\
E_y^r\\
0
\end{pmatrix}.
\end{equation}
Solving the system of equations we obtain
\begin{equation}
\begin{split}
t_x &=\frac{E_x^t}{E_0}=\frac{M_{33}\cos\theta-M_{13}\sin\theta}{M_{11}M_{33}-M_{13}M_{31}} \\
t_y &=\frac{E_y^t}{E_0}=-\frac{M_{31}\cos\theta-M_{11}\sin\theta}{M_{11}M_{33}-M_{13}M_{31}} \\
r_x &=\frac{E_x^r}{E_0}=M_{21}t_x-M_{23}t_y\\
r_y &=\frac{E_y^r}{E_0}=M_{41}t_x-M_{43}t_y.
\end{split}
\end{equation}
From these coefficients, the induced ellipticity per reflection $\varphi$ is calculated with~\cite{Born1983}
\begin{equation}
\sin(2\varphi)=\sin(2\theta)\sin\delta,
\end{equation}
where
\begin{equation}\label{eq:SimuEllip}
\begin{split}
&\tan\theta=\frac{\vert r_y\vert}{\vert r_x\vert}\\
&\delta=\arg\left[\frac{r_y}{r_x}\right],
\end{split}
\end{equation}
where $\delta$ is the phase retardation per reflection of the multilayer. This expression of the ellipticity corrects the one used in~\cite{Bielsa2009}.

\subsection{Results}

As it was first done in~\cite{Bielsa2009}, to reproduce the experimental tendency, we added a anisotropic index of refraction in the layer in contact with the substrate. To reproduce the B-et-al trend we need to assume $n^{(1)}_y=n^{(1)}_x+n_\Delta$ where $n_\Delta$ was 0.13 in~\cite{Bielsa2009} because of an incorrect expression. Using the correct expression Eq.~\ref{eq:SimuEllip}, we set $n_\Delta=0.065$ differing by a factor 2.

To take into account the fact that each layer is birefringent, we add another anisotropy in every layer $\Delta n$ in the $y$ axis of the layers. Knowing that in a layer the induced ellipticity is
\begin{equation}
\varphi^{(i)}=\frac{2\pi}{\lambda}d^{(i)}\Delta n=\frac{\pi}{2n_x^{(i)}}\Delta n,
\end{equation}
for the layer $i$. Because we measured the phase retardation per reflection for two bilayers as $\delta\sim 10^{-7}$~rad, we will assume that the induced ellipticity of one layer is $\varphi^{(1)}=\varphi^{(2)}=2\delta/4=5\times 10^{-8}$~rad.

Since only our work and the one of~\cite{Xiao2019} report the number of layer deposited and since both use a multilayer stack composed of alternating SiO$_2$ and Ta$_2$O$_5$ layers, we use $n^{(1)}_x=2.12$ and $n^{(2)}_x=1.48$~\cite{Xiao2019} as well as $n_s=1.45$. 

\begin{figure}
\centering
\includegraphics[width=\linewidth]{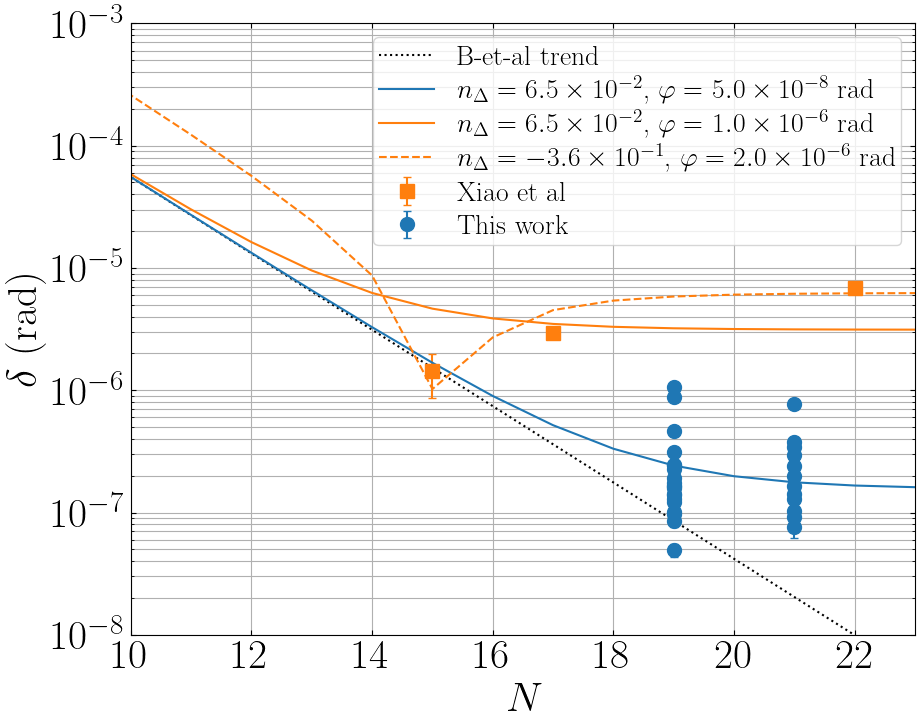}
\caption{Phase retardation per reflection as a function of $N$ the number of deposited bilayers. Data from this work and~\cite{Xiao2019} are reported. The~\cite{Bielsa2009} trend as well as our simulation taking into account layers birefringence are shown}
\label{fig:DeltavsN}
\end{figure}

We finally compare the results reported in this work with our simulation in Fig.~\ref{fig:DeltavsN}. First, we observe that the B-et-al trend cannot reproduce the observed saturation of $\delta$ as a function of $N$, see black dashed line in Fig.~\ref{fig:DeltavsN}. Accounting for the birefringence of each layer we can reproduce the observed global behaviour as shown by the blue curve in Fig.~\ref{fig:DeltavsN} for our data and the orange curve for data~\cite{Xiao2019}. For the~\cite{Xiao2019} data where we see an increase in the phase retardation, they can be explained by a negative birefringence of the first layer, see orange dashed curve in Fig.~\ref{fig:DeltavsN}. Actually, this could also be the case for our data as we observed both an increase and a decrease in phase retardation when adding layers. Facts that in principle our simulation can reproduce. Let us note that in the case of an increase in phase retardation, one need an higher anisotropy in the first layer than for a decrease as shown in Fig.~\ref{fig:DeltavsN} by the trend of the orange solid and dashed line when $N$ tends to 0.

\section{Conclusion}

In this paper, thanks to our capability to deposit additional bilayers while reaching a reflectivity never measured before at $\lambda=1064$~nm of $R=0.99999650$, we have been able to completely characterise the waveplate associated to these bilayers. 

We show that the value of their phase retardation is not random since all measurements give a mean absolute value for $\delta$ of $(1.5\pm 0.4)\times 10^{-7}$~rad. Since it is known that phase retardation can be originated by stress, our measurement seems to confirm this fact in thin films and coatings as discussed in detail in~\cite{Abadias2018}.

As for the angle of the optical axis, they also do not look randomly distributed, the mean value being $\theta_2=(40\pm 4)^\circ$. This measurement question the fabrication processes which are supposed to be isotropic.

Our computational study allows to explain the behaviour of phase retardation as a function of $N$, the number of bilayers, by combining the phase retardation associated by each bilayer with a bigger anisotropy in the first layer in contact with the substrate. Let us note that the stress birefringence that one has to associate to the layer close to the substrate is much higher than the stress measured for a fused silica substrate~\cite{Xiao2018}, for example, and than the stress of each bilayer that we measured, which seems to be unlikely. The presence of the first layer very high anisotropy could be neglected but for data of Xiao~\emph{et~al.} which increase as a function of $N$. Indeed, without their observations, one could reduce the dependence of the phase retardation on $N$, for a large $N$, to a constant corresponding to about the retardation of a bilayer for $N$. The observed B-et-al trend would be left to reflect technological progress in mirror fabrication.

Maybe, an important upgrade of the computational study would be to introduce losses directly in the calculations, with a certain complication of the mathematical formulation. This, to have a simulation closer to reality, and therefore to see if losses can affect the total phase retardation of the multilayer.

As far as the main goal of interest of experiments like BMV to control and eventually erase mirror phase retardation, it looks like the understanding of what gives the favoured direction of the optical axis of a bilayer is the main milestone to be reached.

\section*{Statements and Declarations}
\paragraph{Acknowledgments} We thank all the technical staff of the LNCMI and all the members of the BMV collaboration. We also thank the R\&T board of Safran Electronics \& Defense for funding the realisation of the mirror polishing and deposition. Finally, we deeply thank the teams of the industrial board for allocating machine time, and providing the relevant ressources of the ``process'' team to realise and characterise these mirrors. 

\paragraph{Author contributions} All the authors were involved in the preparation of the manuscript. All the authors have read and approved the final manuscript.

\paragraph{Data Availability Statement} The datasets generated during and/or analyzed during the current study are available from the corresponding author on reasonable request.

\bibliographystyle{epj2}
\bibliography{BiblioAP.bib}

\begin{thebibliography}{38}

\bibitem{Yoshino1979}
T.~Yoshino, {\em Jap. J. Appl. Phys.}, \textbf{18}, 1503 (1979)

\bibitem{Born1983}
M.~Born, E.~Wolf, \emph{Principles of optics}, 6th~edn. (Pergamon Press, 1983)

\bibitem{Millo2009}
J.~Millo, D.V. {Magalh\~aes}, C.~Mandache, Y.~Le~Coq, E.M.L. English, P.G.
  Westergaard, J.~Lodewyck, S.~Bize, P.~Lemonde, G.~Santarelli, {\em Phys. Rev.
  A}, \textbf{79}, 053829 (2009)

\bibitem{Pold2020}
J.H. P{\~{o}}ld, A.D. Spector, {\em {EPJ} Techniques and Instrumentation},
  \textbf{7} (2020)

\bibitem{Martinu2000}
L.~Martinu, D.~Poitras, {\em Journal of Vacuum Science {\&} Technology A},
  \textbf{18}, 2619 (2000)

\bibitem{Bielsa2009}
F.~Bielsa, A.~Dupays, M.~Fouch\'e, R.~Battesti, C.~Robilliard, C.~Rizzo, {\em
  Appl. Phys. B}, \textbf{97}, 457 (2009)

\bibitem{Xiao2019}
S.~Xiao, B.~Li, J.~Wang, {\em Applied Optics}, \textbf{59}, A99 (2019)

\bibitem{Micossi1993}
P.~Micossi, F.~{Della Valle}, E.~Milotti, E.~Zavattini, C.~Rizzo, G.~Ruoso,
  {\em Appl. Phys. B}, \textbf{57}, 95 (1993)

\bibitem{Abadias2018}
G.~Abadias, E.~Chason, J.~Keckes, M.~Sebastiani, G.B. Thompson, E.~Barthel,
  G.L. Doll, C.E. Murray, C.H. Stoessel, L.~Martinu, {\em Journal of Vacuum
  Science {\&} Technology A}, \textbf{36}, 020801 (2018)

\bibitem{Wei2021}
S.~Wei, Y.~Pang, Z.~Bai, Y.~Wang, Z.~Lu, {\em International Journal of Optics},
  \textbf{2021}, 1 (2021)

\bibitem{Ejlli2020}
A.~Ejlli, F.~{Della Valle}, U.~Gastaldi, G.~Messineo, R.~Pengo, G.~Ruoso,
  G.~Zavattini, {\em Physics Reports}, \textbf{871}, 1 (2020)

\bibitem{Beard2021}
J.~B{\'{e}}ard, J.~Agil, R.~Battesti, C.~Rizzo, {\em Review of Scientific
  Instruments}, \textbf{92}, 104710 (2021)

\bibitem{Fan2017}
X.~Fan, S.~Kamioka, T.~Inada, T.~Yamazaki, T.~Namba, S.~Asai, J.~Omachi,
  K.~Yoshioka, M.~Kuwata-Gonokami, A.~Matsuo et~al., {\em The European Physical
  Journal D}, \textbf{71} (2017)

\bibitem{Kryhin2023}
S.~Kryhin, E.D. Hall, V.~Sudhir, {\em Physical Review D}, \textbf{107}, 022001
  (2023)

\bibitem{Agil2022}
J.~Agil, R.~Battesti, C.~Rizzo, {\em The European Physical Journal D},
  \textbf{76}, 192 (2022)

\bibitem{Hartman2017}
M.T. Hartman, A.~Riv{\`{e}}re, R.~Battesti, C.~Rizzo, {\em Review of Scientific
  Instruments}, \textbf{88}, 123114 (2017)

\bibitem{Bouchiat1982}
M.A. Bouchiat, L.~Pottier, {\em Appl. Phys. B}, \textbf{29}, 43 (1982)

\bibitem{Carusotto1989}
S.~Carusotto, E.~Polacco, E.~Iacopini, G.~Stefanini, E.~Zavattini, F.~Scuri,
  {\em Appl. Phys. B}, \textbf{48}, 231 (1989)

\bibitem{Jacob1995}
D.~Jacob, M.~Vallet, F.~Bretenaker, A.L. Floch, M.~Oger, {\em Opt. Lett.},
  \textbf{20}, 671 (1995)

\bibitem{Ni1996}
W.T. Ni, {\em Chin. J. Phys.}, \textbf{34}, 962 (1996)

\bibitem{Wood1996}
C.~Wood, S.C. Bennett, J.L. Roberts, D.~Cho, C.E. Wieman, {\em Opt. Photo.
  News}, \textbf{7}, 54 (1996)

\bibitem{Moriwaki1997}
S.~Moriwaki, H.~Sakaida, T.~Yuzawa, N.~Mio, {\em Appl. Phys. B}, \textbf{65},
  347 (1997)

\bibitem{Brandi1997}
F.~Brandi, F.~{Della Valle}, A.M. De~Riva, P.~Micossi, F.~Perrone, C.~Rizzo,
  G.~Ruoso, G.~Zavattini, {\em Appl. Phys. B}, \textbf{65}, 351 (1997)

\bibitem{Hall2000}
J.L. Hall, J.~Ye, L.S. Ma, {\em Phys. Rev. A}, \textbf{62}, 013815 (2000)

\bibitem{Lee2000}
J.Y. Lee, H.W. Lee, J.W. Kim, Y.S. Yoo, J.W. Hahn, {\em Appl. Opt.},
  \textbf{39}, 1941 (2000)

\bibitem{Morville2002}
J.~Morville, D.~Romanini, {\em Appl. Phys. B}, \textbf{74}, 495 (2002),
  10.1007/s003400200854

\bibitem{Huang2008}
H.~Huang, K.K. Lehmann, {\em Appl. Opt.}, \textbf{47}, 3817 (2008)

\bibitem{Durand2009}
M.~Durand, Ph.D. thesis, Universit\'e Claude Bernard, Lyon 1 (2009),
  \url{https://tel.archives-ouvertes.fr/tel-00432201}

\bibitem{Cadene2015}
A.~Cad\`ene, Ph.D. thesis, Universit\'e Toulouse 3 Paul Sabatier (2015),
  \url{http://thesesups.ups-tlse.fr/2743/1/2015TOU30037.pdf}

\bibitem{DellaValle2016}
F.~{Della Valle}, A.~Ejlli, U.~Gastaldi, G.~Messineo, E.~Milotti, R.~Pengo,
  G.~Ruoso, G.~Zavattini, {\em The European Physical Journal C}, \textbf{76}
  (2016)

\bibitem{Rivere2017}
A.~Riv\`ere, Ph.D. thesis, Universit\'e Toulouse 3 Paul Sabatier (2017),
  \url{http://thesesups.ups-tlse.fr/4187/1/2017TOU30387.pdf}

\bibitem{Kamioka2020}
S.~Kamioka, Ph.D. thesis, Department of Physics, Graduate School of Science,
  The University of Tokyo (2020), \url{https://tabletop.icepp.s.u-tokyo.ac.
  jp/wp-content/uploads/2021/02/Dron-kamioka.pdf}

\bibitem{JCGM2008}
JCGM, \emph{Evaluation of measurement data : Guide to the expression of
  uncertainty in measurement} (2008)

\bibitem{Drever1983}
R.W.P. Drever, J.L. Hall, F.V. Kowalski, J.~Hough, G.M. Ford, A.J. Munley,
  H.~Ward, {\em Appl. Phys. B}, \textbf{31}, 97 (1983)

\bibitem{Bielsa2005}
F.~Bielsa, R.~Battesti, C.~Robilliard, G.~Bialolenker, G.~Bailly, G.~Tr\'enec,
  A.~Rizzo, C.~Rizzo, {\em Eur. Phys. J. D}, \textbf{36}, 261 (2005)

\bibitem{DellaValle2014}
F.~{Della Valle}, E.~Milotti, A.~Ejlli, U.~Gastaldi, G.~Messineo,
  L.~Piemontese, G.~Zavattini, R.~Pengo, G.~Ruoso, {\em Opt. Express},
  \textbf{22}, 11570 (2014)

\bibitem{Rempe1992}
G.~Rempe, R.~Lalezari, R.J. Thompson, H.J. Kimble, {\em Opt. Lett.},
  \textbf{17}, 363 (1992)

\bibitem{Xiao2018}
S.~Xiao, B.~Li, H.~Cui, J.~Wang, {\em Optics Letters}, \textbf{43}, 843 (2018)

\end{thebibliography}

\end{document}